\documentclass[aps,twocolumn,prb,preprintnumbers,amsmath,amssymb,superscriptaddress,floats]{revtex4-2}

\usepackage[utf8]{inputenc}
\usepackage{amsfonts}
\usepackage{amsmath}
\usepackage{amssymb}
\usepackage{bbm}
\usepackage{amsthm}
\usepackage{empheq}
\usepackage{enumitem}
\usepackage{verbatim}
\usepackage[dvips]{color}
\usepackage{subfigure}
\usepackage[breaklinks]{hyperref}
\usepackage{graphicx}
\usepackage{siunitx}
\usepackage{times}

\begin{document}

\title{Weyl-Luttinger phase transition in pyrochlore iridates revealed by Raman scattering}
\author{Predrag Nikoli\'c}
\affiliation{Department of Physics and Astronomy, George Mason University, Fairfax, VA 22030, USA}
\affiliation{Institute for Quantum Matter at Johns Hopkins University, Baltimore, MD 21218, USA}
\author{Yuanyuan Xu}
\affiliation{Institute for Quantum Matter at Johns Hopkins University, Baltimore, MD 21218, USA}
\author{Takumi Ohtsuki}
\affiliation{Institute for Solid State Physics, University of Tokyo, Kashiwa, Chiba 277-8581, Japan}
\author{Satoru Nakatsuji}
\affiliation{Institute for Quantum Matter at Johns Hopkins University, Baltimore, MD 21218, USA}
\affiliation{Department of Physics, University of Tokyo, Bunkyo-ku, Tokyo 113-0033, Japan}
\affiliation{Institute for Solid State Physics, University of Tokyo, Kashiwa, Chiba 277-8581, Japan}
\affiliation{CREST, Japan Science and Technology Agency, Kawaguchi, Saitama 332-0012, Japan}
\affiliation{Trans-scale Quantum Science Institute, University of Tokyo, Bunkyo-ku, Tokyo 113-0033, Japan}
\author{Natalia Drichko}
\affiliation{Institute for Quantum Matter at Johns Hopkins University, Baltimore, MD 21218, USA}

\date{\today}

\begin{abstract}

Pyrochlore iridates are thought to be a fertile ground for the realization of topologically non-trivial and strongly correlated states of matter. Here we observe a magnetically-driven phase transition between the topological Weyl semimetal and a Luttinger semimetal with quadratic band touching in Nd$_2$Ir$_2$O$_7$ using Raman scattering. We also find evidence of quadratic band touching in Pr$_2$Ir$_2$O$_7$, with the chemical potential further away from the node. Our theoretical analysis of the Raman scattering from Weyl and Luttinger quasiparticles agrees with experimental observations, and enables a characterization of the material parameters while revealing interaction and disorder effects through a relatively short quasiparticle lifetime. 

\end{abstract}

\maketitle

Strongly correlated electrons with topologically non-trivial dynamics are among the most interesting physical systems that host unconventional states of matter \cite{Nagaosa2013, Fert2013, Witczak2014, Takagi2019, Broholm2020, nakatsuji_annurev-conmatphys-031620-103859}. The extreme topological interaction effect, electron fractionalization \cite{WenQFT2004, senthil00, Nikolic2019}, is sought for quantum computing applications \cite{Bonderson2010}, while the more ubiquitous ability of interactions to shape phase transitions \cite{SubirQPT} is a means to obtain new transport properties and control materials in devices \cite{nakatsuji_annurev-conmatphys-031620-103859}. Recent extensive studies have uncovered prominent realizations of magnetic Weyl semimetals that attract much attention in a vast range of fields, from basic science to applications \cite{Armitage2018, nakatsuji_annurev-conmatphys-031620-103859, Nakatsuji2015, Sakai2018, Liu2018, Liu2019a, Belopolski2019}. Among topological magnets, pyrochlore iridates are a promising material family expected to host Weyl \cite{Machida2010, Ari2010, Goswami2016a, Roy2021, Ohtsuki2019} and Luttinger semimetal states \cite{Kondo2015, Cheng2017}, as well as correlated topological insulator states \cite{Pesin2010, Tian2016}. Theoretically, these states universally emerge from a non-Fermi liquid quantum critical point \cite{Nakatsuji2006, Moon2013, Krempa2013, Tokiwa2014, Herbut2014, Kondo2015, Krempa2019, Boettcher2019, Polini2019, Herbut2020, Mandal2021b, Freire2021} with quadratic band touching, which was predicted long ago \cite{Abrikosov1971}.

The experimental exploration of the topological correlation physics in pyrochlore iridates is gaining momentum, but the confirmation and characterization of topologically non-trivial electron bands is still a challenge \cite{Machida2010, Kondo2015, Nakayama2016, Tian2016, Cheng2017, Armitage2020}. In Nd$_2$Ir$_2$O$_7$, where the higher-temperature paramagnetic phase is expected to host a quadratic band touching, Weyl electrons would arise below $T_\mathrm{N} = 33$~K due to the time-reversal symmetry breaking associated with the antiferromagnetic order of Ir moments. This magnetic semimetal would persist down to $T=14$ K when the system enters a fully gapped semiconducting phase due to the ordering of Nd moments \cite{Tomiyasu2012, Guo2013, Tian2016, Armitage2020}. However, evidence for Weyl fermions could not be obtained with ARPES due to the low energy resolution \cite{Nakayama2016}, and has been indicated with optical spectroscopy (although, possessing unconventional properties caused by interactions) \cite{Armitage2020}. In this work, we report the electronic contribution to Raman scattering in Nd$_2$Ir$_2$O$_7$ and Pr$_2$Ir$_2$O$_7$, and also calculate the Raman response of Weyl and Luttinger nodal electrons using perturbation theory. By matching the measured frequency dependence of the Raman intensity with theory, we reveal the presence of Weyl quasiparticles in the low-temperature phase of Nd$_2$Ir$_2$O$_7$, as well as Luttinger quasiparticles in Pr$_2$Ir$_2$O$_7$ and the high-temperature phase of Nd$_2$Ir$_2$O$_7$. The magnetic ordering in Nd$_2$Ir$_2$O$_7$ at $T_\mathrm{N} = 33$~K was detected by our Raman scattering experiment through the observation of magnons and will be discussed elsewhere \cite{Xu2022prep}. In this study, we finally reveal that this is also a topological phase transition associated with the splitting of the Luttinger quadratic band-touching node into a set of Weyl points, driven by the time-reversal symmetry breaking. The theory-experiment comparison allows us to also estimate the Fermi energy, quasiparticle lifetime and other properties of the Weyl spectrum in Nd$_2$Ir$_2$O$_7$.

Electrons contribute an incoherent background to the Raman differential scattering cross-section $\partial^{2}\sigma / \partial\Omega\partial\omega_{\textrm{s}} = (\omega_{\textrm{s}}/\omega_{\textrm{i}}) (e^{2}/mc^{2})^{2} \chi''(\Omega)$ via two-photon processes in which a virtual particle-hole excitation is created by absorbing an incoming photon of frequency $\omega_{\textrm{i}}$ and quickly dissolved by emitting a scattered photon of frequency $\omega_{\textrm{s}}$. The measured dependence of $\chi''(\Omega)$ on the energy transfer $\Omega=\omega_{\textrm{i}}-\omega_{\textrm{s}}$ to the electron gas (in the units $\hbar=1$), known as Raman shift frequency, provides rich information about the electron dynamics. If the quasiparticles with conduction and valence bands have an infinite lifetime, then $\chi''(\Omega)$ directly reflects the electronic density of states above a threshold frequency set by the chemical potential $\mu$. The Raman amplitude $\chi''(\Omega)=-\frac{1}{\pi} \textrm{Im} \lbrace \chi(\Omega, {\bf q}\to0) \rbrace$ is extracted from a response function
\begin{equation}\label{Chi1}
\chi({\bf q},\Omega) = -i\sum_{{\bf k}\omega}\textrm{tr}\Bigl\lbrace G({\bf k},\omega)\gamma_{{\bf k}+{\bf q}}G({\bf k}+{\bf q},\omega+\Omega)\gamma_{\bf k}\Bigr\rbrace \ ,
\end{equation}
where $G({\bf k},\omega)$ is the electron Green's function and $\gamma_{\bf k}$ is the Raman vertex function \cite{Deveraux2007}. In order to capture the Weyl spectrum and basic interaction effects, we adopt a simple model Green's function
\begin{equation}\label{Green2}
G({\bf k},\omega)=\frac{A({\bf k},\omega)}{\omega - \left(\frac{k^2}{2m} + v\boldsymbol{\sigma}{\bf k} - \mu \right) + i\Gamma(\omega) \textrm{sign}(\omega)} \ .
\end{equation}
A quadratic term involving an effective mass $m$ is retained in order to utilize the existing non-relativistic theory of Raman scattering \cite{Deveraux2007} while producing a more realistic spectrum. The pure relativistic limit is achieved simply by setting $m\to\infty$. The spin-orbit coupling involves the vector $\boldsymbol{\sigma}$ of Pauli matrices and gives the Green's function a matrix character. Since the Weyl Hamiltonian can be effectively expressed as a Hamiltonian $({\bf k}-\boldsymbol{\mathcal{A}})^2/2m$ of ordinary electrons coupled to an SU(2) gauge field $\boldsymbol{\mathcal{A}}=-mv\boldsymbol{\sigma}$, the spin-orbit coupling is incorporated into the non-relativistic Raman vertex merely by a momentum shift ${\bf k} \to {\bf k} - \boldsymbol{\mathcal{A}}$. A finite quasiparticle lifetime $\tau$ is captured by the imaginary part $\Gamma(\omega)\propto\tau^{-1}$ of the electron self-energy correction. Coulomb interactions do not substantially alter the linear Weyl spectrum, so the electron dispersion assumed in the Green's function readily includes renormalizations inherited from the real part of the self-energy.

In case the chemical potential $\mu$ is shifted away from the exact position at the Weyl node, both intraband and interband transitions forming particle-hole pairs contribute to the electronic Raman amplitude $\chi''(\Omega)$. The former contributes \cite{Deveraux2007}
\begin{equation}\label{ChiMetal}
\chi''(\Omega) \propto \frac{\tau\Omega}{1+(\tau\Omega)^2}
\end{equation}
and vanishes as in ordinary metals when the quasiparticle lifetime $\tau$ is infinite. The calculated interband part is shown in Fig.\ref{WLinterband}(a) for the Weyl nodes whose spectrum has spherical symmetry (see the Supplementary Material (SM) \footnote{\label{SM} See Supplemental Material at [URL will be inserted by publisher]} for details). The essential structure of the Raman response is best appreciated in the idealized case of non-interacting Weyl electrons which have infinite lifetime, shown by the red discontinuous curve:
\begin{equation}
\chi''(\Omega) = \frac{m^{2}v\Omega^{2}}{2(2\pi)^{3}}\;\theta\left(\frac{|\Omega|}{2}-\left\vert \mu-\frac{\Omega^{2}}{8mv^{2}}\right\vert \right)\; I(\hat{{\bf e}}_{\textrm{s}},\hat{{\bf e}}_{\textrm{i}}) \ .
\end{equation}
In the relativistic limit $m\to\infty$ (well-defined only for the measured differential scattering cross-section), the response occurs only above the Pauli threshold frequency $\Omega>|2\mu|$ for the generation of particle-hole excitations, where the chemical potential $\mu$ is expressed relative to the node energy. Above the threshold, the response follows the density of states $\chi''(\Omega)\propto \Omega^2$ of Weyl electrons. Realistically, however, interactions and disorder impart a finite lifetime on Weyl quasiparticles without qualitatively affecting the energy spectrum. A finite lifetime is seen to quickly fill-in the Raman response below the threshold with a linear-looking frequency dependence. The analogous calculation of the Raman response in an idealized Luttinger semimetal \cite{Moon2013}, whose quasiparticles exhibit a quadratic band touching, produces $\chi''(\Omega) \propto \sqrt{\Omega}$ above the threshold $\Omega>|2\mu|$ which is characteristic for the density of states from a quadratic energy dispersion. The effect of a finite Luttinger quasiparticle lifetime on interband Raman transitions is shown in Fig.\ref{WLinterband}(b) (see SM \cite{Note1} for details).

\begin{figure}
\includegraphics[width=0.45\textwidth]{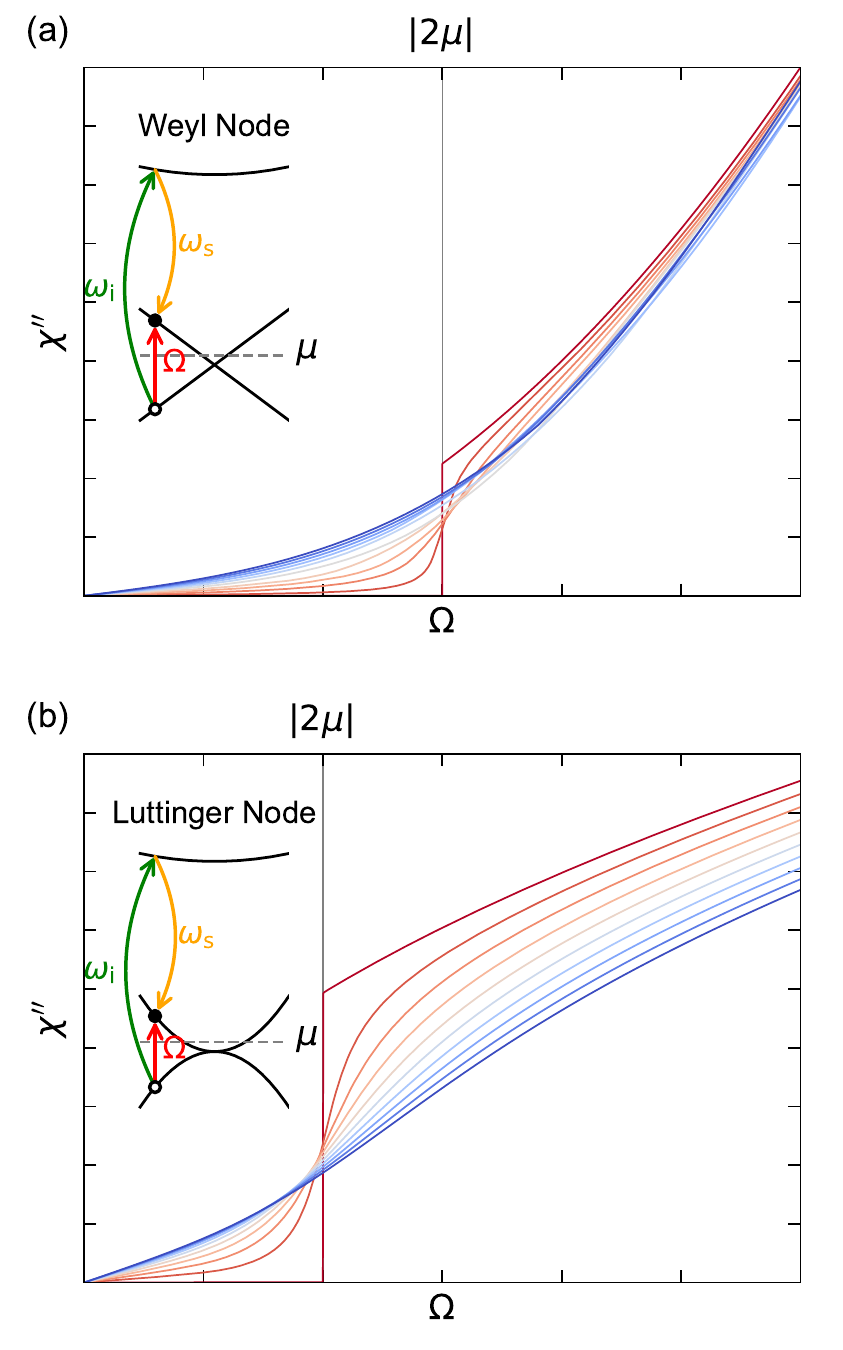}
\caption{Frequency dependence of the Raman intensity from interband transitions of (a) Weyl and (b) Luttinger nodal quasiparticles. The vertical line indicates a threshold frequency $\Omega=|2\mu|$, below which the Raman scattering vanishes when the quasiparticles have infinite lifetime. The outermost red curve represents the infinite lifetime: it sharply drops to zero at $\Omega<|2\mu|$, and otherwise follows $\chi''(\Omega)\propto\Omega^2$ in (a) and $\chi''(\Omega)\propto\sqrt{\Omega}$ in (b). The other smooth curves are obtained with the inverse lifetime $\Gamma\propto\tau^{-1}$ increasing in steps of 0.1, up to 0.9 (expressed in the same arbitrary units as $\Omega$).}
\label{WLinterband}
\end{figure}

We compare our theoretical results on the Raman scattering response of the Luttinger and Weyl nodes to the experimental electronic Raman scattering spectra of Nd$_2$Ir$_2$O$_7$ and Pr$_2$Ir$_2$O$_7$. Earlier studies have suggested that the  cubic symmetry should protect the existence of Luttinger nodal electrons with quadratic band touching in the paramagnetic state \cite{Abrikosov1971, Moon2013}. Applied to these particular pyrochlore iridates, both numerous band structure calculations \cite{Pesin2010, Chen2012, Ishii2015, Shinaoka2019, Zhang2017, Wang2017} and some experimental results \cite{Kondo2015, Nakayama2016, Cheng2017} point to the Luttinger node in the band structure of a paramagnetic state. Nd$_2$Ir$_2$O$_7$ exhibits a phase transition associated with magnetic ordering of Ir moments at $T_\mathrm{N} = 33$ K \cite{Armitage2020,Xu2022prep}. The broken time-reversal symmetry (TRS) in the low-temperature phase with ordered Ir moments is expected to split the Luttinger node into a set of Weyl nodes \cite{Ari2010}. We observe an evidence of this topological phase transition with Raman scattering.

\begin{figure}
\includegraphics[width=0.45\textwidth]{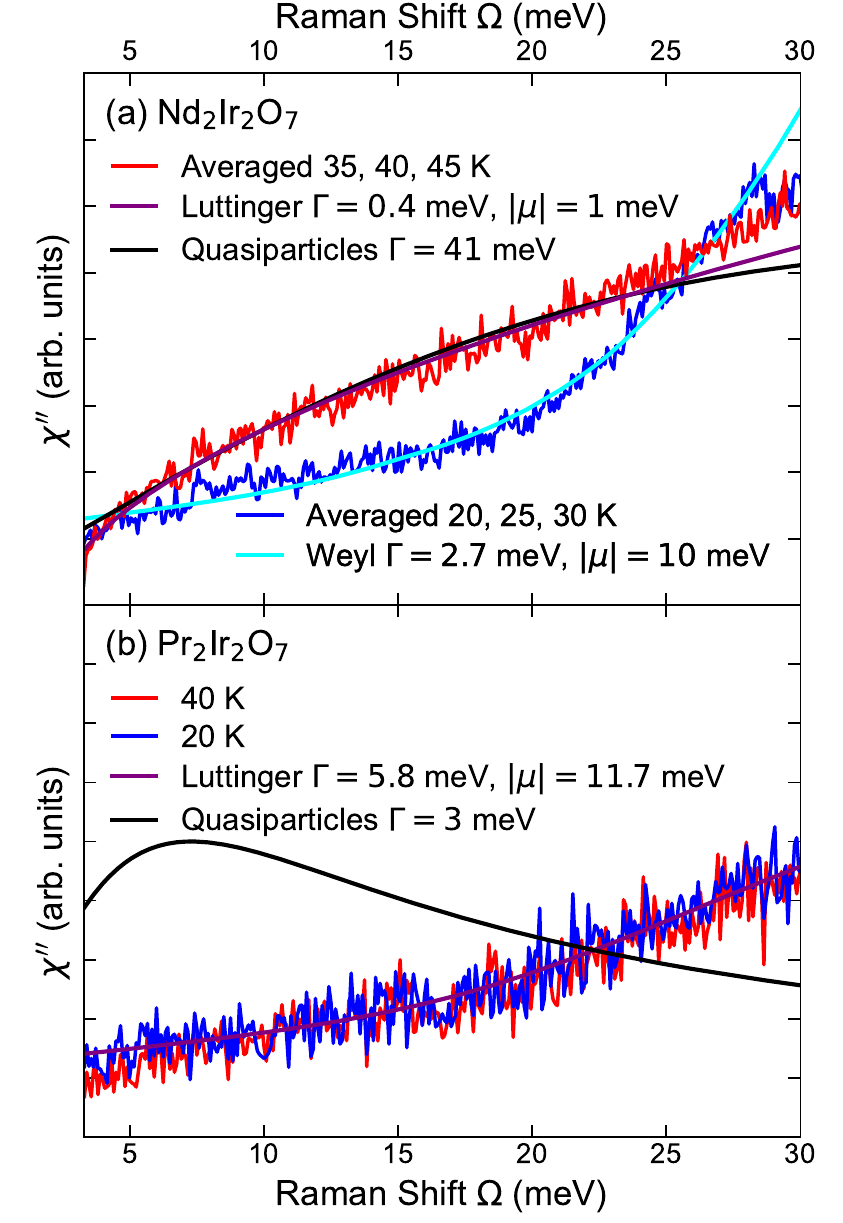}
\caption{\label{RamanWeyl}Upper panel: Raman scattering spectra of Nd$_2$Ir$_2$O$_7$ above (red) and below (blue) the ordering temperature $T_N=33$ K of Ir moments. The high-temperature spectrum is compared with the behavior of finite-lifetime charge carriers (black line), and interband transitions in the case of quadratic band touching (purple line). The low temperature Raman spectrum is well described only by interband transitions of a Weyl semimetal (cyan line). Assuming that all Weyl nodes contributing to the Raman signal have the same node energy, we estimate the chemical potential $|\mu| \approx 10$ meV relative to the node. Lower panel: Raman scattering spectra of Pr$_2$Ir$_2$O$_7$ in the temperature range $35-45$ K (red), and $20-30$ K (blue). These spectra are compared with the behavior of finite-lifetime charge carriers using the $\Gamma$ values from Ref.~\cite{Armitage2018} (black line), and interband transitions in the case of quadratic band touching where disorder and interactions give a linear-looking frequency dependence (purple line). Note that the interband transition fits in both panels do not include small charge carrier (Fermi surface) contributions (\ref{ChiMetal}), but these contributions are visible in the low-frequency data as small deviations from the fits.}
\end{figure}

The Nd$_2$Ir$_2$O$_7$ and Pr$_2$Ir$_2$O$_7$ Raman scattering spectra above $30$ meV consist of sharp phonon modes \cite{Xu2022} and magnon bands for Nd$_2$Ir$_2$O$_7$ in the low temperature state \cite{Xu2022prep}, superimposed on a weaker broad electronic continuum. Raman response $\chi''(\Omega)$ in the $(x,y)$ scattering channel contains $E_g$ and $T_{2g}$ components of scattering.  We obtained the electronic contribution to the Raman scattering spectra by subtracting the phonon and magnon contributions from the Raman signal. The upper panel of Fig.\ref{RamanWeyl}  compares the experimental electronic Raman scattering spectrum of Nd$_2$Ir$_2$O$_7$ in the high-temperature paramagnetic ($T\sim35-45$ K) and low-temperature magnetic ($T\sim20-30$ K) phases. We observe a considerable difference between the frequency dependence of the electronic Raman continuum above and below the phase transition temperature, which can be well understood using our theoretical insight into the $\chi''(\Omega)$ response for Luttinger and Weyl nodes.

The Raman scattering of nodal electrons is dominated by interband transitions when the Fermi energy $\mu$ is close to the node energy. Such interband processes produce the Raman amplitude shown in Fig.\ref{WLinterband}. Alternatively, when the Fermi energy is shifted considerably from the band touching energy, the ample charge carriers available in the ground state may dominate the Raman response at low frequencies and give rise to a conventional Raman signal (\ref{ChiMetal}) found in ordinary metals \cite{Deveraux2007} and superconductors \cite{Hackl1996}. The two types of Raman response are qualitatively different, so a comparison of the measured electronic Raman response to the theoretical predictions can give us an insight into the origin of the electronic scattering.

We compare the electronic Raman scattering spectra of Nd$_2$Ir$_2$O$_7$ to the theoretical models in Fig.~\ref{RamanWeyl}(a). The paramagnetic phase in the temperature range $T\sim30-45$ exhibits a Raman spectrum which can be naively fitted by two models: intraband transitions (\ref{ChiMetal}) of conventional charge carriers (CCC, black line), and interband transitions of Luttinger nodal quasiparticles (LNQ, purple line). However, the model CCC fails in several ways. First of all, LNQ provides a better fit in a wider frequency range, specifically above $\Omega>20$ meV (the CCC curve eventually turns down and decreases with growing frequency). More importantly, the parameters which produce a good CCC fit, i.e. the relaxation rate $\Gamma\propto\tau^{-1}=41$~meV of free charge carriers at $T\sim35-45$ K, do not seem physical \footnote{Exact formula for the fit: $\chi''(\omega) = 0.059 + 0.25 \tau\omega/\lbrack 1+(\tau\omega)^2\rbrack$, with $\tau^{-1}=41$ ~meV for 35-45 K averaged spectra}. The THz measurements performed on the related Pr$_2$Ir$_2$O$_7$ compound are consistent with considerably smaller Drude peak values, $\Gamma\approx$ 8.5~meV in the $T\sim30-50$ K range \cite{Armitage2017b}, similar to the values for Nd$_2$Ir$_2$O$_7$ \cite{Armitage2020}.

A striking indication of nodal interband transitions is found in the magnetic phase at $T < T_\mathrm{N}$. The Raman amplitude drops abruptly at $T < T_\mathrm{N}$ across a wide frequency range, but picks up and grows at higher frequencies. In this temperature range, the Raman spectrum is described best by the interband transitions of Weyl electrons with a finite lifetime.  In Fig.~\ref{RamanWeyl}(a), we compare the experimental spectra to the theoretical curve of $\chi''(\Omega)$ for Weyl nodes with $\Gamma=2.7$ meV and $|\mu|=10$ meV, which produces a good agreement up to  about $\Omega\sim 30$ meV which could be the energy cut-off for the linear Weyl spectrum. The quantity $\chi''(\Omega)$ is not very sensitive to small changes of parameters, so $|\mu|=10$ meV in the magnetic phase was selected to be close to the cut-off associated with the charge carriers' contribution to the IR spectra   of Nd$_2$Ir$_2$O$_7$ at these temperatures \cite{Armitage2020}. 

\begin{figure}
\centering
\includegraphics[width=0.45\textwidth]{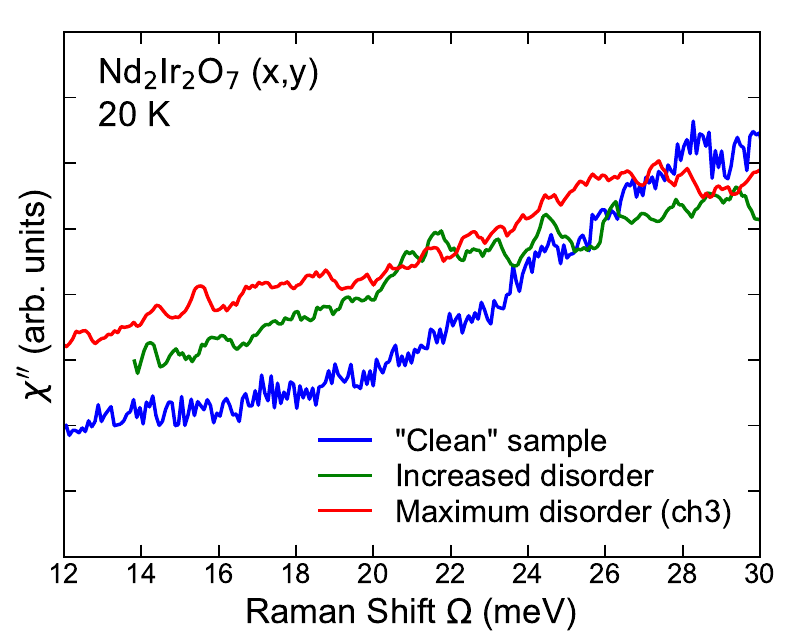}
\caption{Raman scattering spectra of three samples of Nd$_2$Ir$_2$O$_7$ with different amounts of disorder at $T=20$~K, which is  below $T_\mathrm{N}$.  Note the increase of Raman intensity compared to the response of the ``clean'' sample presented in Fig.~\ref{RamanWeyl} and the linear dependence on frequency for the samples with disorder. Structural disorder was characterized by resistivity measurements and phonon widths (see SM \cite{Note1}). }
\label{nio_disorder}
\end{figure}

Fig~\ref{nio_disorder} compares the  $T=20$~K electronic Raman scattering spectra of three different samples of Nd$_2$Ir$_2$O$_7$ with increasing structural disorder (see SM \cite{Note1} for the characterization of structural disorder). The results for the sample with the least disorder at $T > T_N$ and $T < T_N$ are presented in Fig.~\ref{RamanWeyl}(a). Disorder reduces the lifetime of excitations, and could shift the chemical potential. We observe a change of the electronic scattering with increasing disorder, where the low-frequency contribution to $\chi''(\Omega)$ becomes linear and increases in intensity, masking the $\chi''(\Omega) \propto \Omega^2$ behaviour, as predicted by the calculations. 

An example of a quadratic node which is preserved on cooling between 40 and 20~K is provided by Pr$_2$Ir$_2$O$_7$.  Electronic band structure of this material is similar to the paramagnetic state of Nd$_2$Ir$_2$O$_7$, but  with $\mu$ shifted further away from the nodal position. Studies of Pr$_2$Ir$_2$O$_7$ films suggest the shift up to  17 meV  \cite{Armitage2017b}. There is an evidence of Pr$_2$Ir$_2$O$_7$ crystals showing disordered electronic potential \cite{Kavai2021}.  In Fig~\ref{RamanWeyl}(b)  we present experimental electronic Raman scattering data for Pr$_2$Ir$_2$O$_7$, which show the same response for 40 and 20~K, in accordance with the absence of changes of magnetic or electronic structure in this temperature range. A comparison with the conventional response of charge carries (Eq.~\ref{ChiMetal}), using $\Gamma=3$ meV at 20~K from the THz spectroscopy studies\cite{Armitage2017b} (black curve), shows that the electronic Raman response is determined by the transitions within the quadratic node. The linear dependence of experimental $\chi''(\Omega)$ up to about 25~meV in the spectra of  Pr$_2$Ir$_2$O$_7$ corresponds to the situation with a high chemical potential ($|\mu|>$ 11 meV) and low lifetime of excitations due to the disorder expected for this material. We have to note here that the decrease of the quasiparticle lifetime, which leads to the linear frequency dependence of the Raman scattering continuum, reduces the distinctions between the Luttinger and Weyl nodes' response.


In conclusion, the electronic Raman scattering of Nd$_2$Ir$_2$O$_7$ agrees well with the calculation of Raman response by interband excitations. The incoherent $\chi''\sim\Omega^2$ response from Weyl nodes at $T<T_N$ and $\chi''\sim\sqrt{\Omega}$ response from Luttinger nodes at $T>T_N$ are not easy to obtain by alternative mechanisms. Specifically, the interband response of Weyl electrons cannot be confused with intraband transitions across a Fermi surface. Instead, some contamination of the response could come from the transitions involving higher-energy states within the Weyl spectrum where the electron dispersion becomes non-linear. Generally, a lattice Weyl-electron model will produce a non-linear low-energy spectrum $E_{\bf k} \sim vk + \alpha k^l$. The sub-dominant power $l>1$, treated using the effective mass approximation \cite{Deveraux2007}, gives $\chi''(\Omega)\sim |\Omega|^{2(l-1)+d-3}$ beyond the threshold ($d=3$ describes the Weyl materials and $d=2$ describes graphene). The $l=2$ power, contributing to $\chi''\sim\Omega^2$, was actually taken into account here through the mass parameter $m$ (without relying on the effective mass approximation). Other powers $l\ge 3$ produce qualitatively different Raman spectra which are not seen in the experiment. Note that virtual transitions to higher-energy bands cannot contaminate the Raman response of Weyl electrons because the effective mass approximation, well-justified in this case, predicts a vanishing Raman vertex $\gamma_{\bf k}$, and hence $\chi''=0$, for the linear energy spectrum.

The interband and intraband responses of electrons with quadratic band touching are notably harder to distinguish, but in Nd$_2$Ir$_2$O$_7$ we presented two arguments in favor of the interband transitions. Another argument can be based on the observation in Fig.\ref{RamanWeyl} that the electronic density of states clearly migrates to lower energies upon the transition to the paramagnetic phase. This must drag the chemical potential $\mu$ toward the node energy. Since $\mu$ was already close to the node energy in the magnetic Weyl-node phase, it can only get closer to the node in the Luttinger-node phase and hence promote the interband Raman transitions. The migration of $\mu$ to a lower energy contributes to the fact that no apparent threshold frequency is seen in the Raman spectrum of the paramagnetic phase.

A Raman experiment alone can reliably distinguish between the true nodal spectrum and the presence of a small bandgap only if electrons have a long lifetime and the Fermi energy sits exactly at the putative node energy. Otherwise, the judgement in favor of nodes must rely on theoretical node protection mechanisms: topological in the case of a linear Weyl spectrum \cite{Ari2010}, and cubic symmetry-based in the case of a Luttinger quadratic band touching \cite{Abrikosov1971}.

The quadratic nodal spectrum is additionally important as a non-Fermi liquid quantum critical point strongly affected by Coulomb interactions \cite{Nakatsuji2006, Moon2013, Tokiwa2014, Cheng2017}. It has been predicted \cite{Freire2022} that the Raman response $\chi''\sim|\Omega|^{\frac{1}{2}+\delta}$ in this state acquires a small correction $\delta\approx 9/76$. This is difficult to experimentally discern without a pristinely accurate fit at low energies. Further studies are needed in order to determine if the Fermi energy in quadratic band touching scenarios indeed sits close enough to the nodes \cite{Onoda2015, Kondo2015} to activate the non-Fermi liquid physics. Looking toward the future, it is highly important to reveal the characters of the non-Fermi-liquid excitations by tuning the Fermi level to the node.

\section{Acknowledgements}

This work was supported at the Institute for Quantum Matter, an Energy Frontier Research Center funded by the U.S. Department of Energy, Office of Science, Basic Energy Sciences under Award No. DE-SC0019331. This work was also partially supported by CREST (JPMJCR18T3), Japan Science and Technology Agency, and by Grants-in-Aid for Scientific Research (19H00650) from the Japanese Society for the Promotion of Science (JSPS).


%

\clearpage
\newpage

\renewcommand\thefigure{S\arabic{figure}}
\setcounter{figure}{0}

\section{Supplementary Material}

\subsection{Electronic Raman scattering cross-section from interband transitions}

The matrix elements of the electronic Raman vertex function $\gamma_{\bf k}$ in the main text are given by \cite{Deveraux2007}:
\begin{eqnarray}\label{Gamma1}
&& \gamma_{\alpha\beta}^{\textrm{i,s}} = \langle\alpha| \gamma_{\bf k}^{\phantom{x}} |\beta\rangle = \hat{{\bf e}}_{\textrm{i}}^{\phantom{x}}\hat{{\bf e}}_{s}^{\phantom{x}}\rho_{\alpha\beta}({\bf q}_{\textrm{i}}-{\bf q}_{\textrm{s}}) +\frac{1}{m}\sum_{\gamma} \\
&& ~ \times\!\left\lbrack \frac{p_{\alpha\gamma}^{\phantom{x}}(\hat{{\bf e}}_{\textrm{s}}^{\phantom{\dagger}},-{\bf q}_{\textrm{s}}^{\phantom{\dagger}})p_{\gamma\beta}^{\phantom{x}}(\hat{{\bf e}}_{\textrm{i}}^{\phantom{\dagger}},{\bf q}_{\textrm{i}}^{\phantom{\dagger}})}{E_{\beta}-E_{\gamma}+\omega_{\textrm{i}}}+\frac{p_{\alpha\gamma}^{\phantom{x}}(\hat{{\bf e}}_{\textrm{i}}^{\phantom{\dagger}},{\bf q}_{\textrm{i}}^{\phantom{\dagger}})p_{\gamma\beta}^{\phantom{x}}(\hat{{\bf e}}_{\textrm{s}}^{\phantom{\dagger}},-{\bf q}_{\textrm{s}}^{\phantom{\dagger}})}{E_{\beta}-E_{\gamma}-\omega_{\textrm{s}}}\right\rbrack \nonumber
\end{eqnarray}
where
\begin{eqnarray}
\rho_{\alpha\beta}^{\phantom{x}}({\bf q}) = \rho_{\beta\alpha}^{*}(-{\bf q}) &=& \langle\alpha|e^{i{\bf q}{\bf r}}|\beta\rangle \\
p_{\alpha\beta}^{\phantom{x}}(\hat{{\bf e}},{\bf q}) = p_{\beta\alpha}^{*}(\hat{{\bf e}},-{\bf q}) &=& \hat{{\bf e}}\,\langle\alpha|e^{i{\bf q}{\bf r}}(-i\boldsymbol{\nabla}-\sigma^{a}{\bf A}^{a})|\beta\rangle \nonumber \ .
\end{eqnarray}
The quasiparticle's initial $|\beta\rangle$ and final $|\alpha\rangle$ states have momenta ${\bf k}$ and ${\bf k}+{\bf q}$ respectively, with ${\bf q}$ being the momentum transfer from the photon to the electrons. The polarizations of the incoming (i) and scattered (s) photon are given by $\hat{{\bf e}}_{\textrm{i}}$ and $\hat{{\bf e}}_{\textrm{s}}$ respectively. The SU(2) gauge field $\boldsymbol{\mathcal{A}} = {\bf A}^a \sigma^a = -mv \boldsymbol{\sigma}$ embodies the spin-orbit coupling which produces the Weyl spectrum, as explained in the main text.

The electron Green's function shaped by disorder and interactions is modelled as
\begin{equation}\label{Green2}
G({\bf k},\omega)=\frac{A({\bf k},\omega)}{\omega-H_{{\bf k}}+i\Gamma({\bf k},\omega) \, \textrm{sign}(\omega)} \ .
\end{equation}
For simplicity, we ignore the spin, frequency and momentum dependence of the spectral weight $A_\sigma({\bf k},\omega) \approx 1$ and the imaginary self-energy part $\Gamma({\bf k},\omega) \approx \Gamma>0$. Despite being very crude, this approximation  faithfully retains the essential effects of all processes which impart a finite lifetime $\tau \propto \Gamma^{-1}$ on the electronic quasiparticles. The matrix representation $H_{\bf k}$ of the single-particle Hamiltonian depends on the type of the nodal spectrum.

\subsection{Weyl spectrum}

The spherically symmetric Weyl spectrum obtains from $H_{\bf k} = k^2/2m + v\boldsymbol{\sigma}{\bf k}$. The mass parameter $m\neq 0$ allows us to both apply the existing non-relativistic theory of Raman scattering and extract the relativistic behavior by taking the $m\to\infty$ limit at the end. In the representation which diagonalizes $H_{\bf k}$, the interband electron scattering contributes
\begin{equation}
\chi({\bf q}\to0,\Omega) = -i\int\frac{d^{3}k}{(2\pi)^{3}}\sum_{\sigma}\gamma_{\sigma,-\sigma}^{\textrm{i,s}}({\bf k})\,\gamma_{-\sigma,\sigma}^{\textrm{s,i}}({\bf k})\, I_{\sigma}^{\phantom{x}}(\Omega) \ , \nonumber
\end{equation}
to the Raman response function, where
\begin{eqnarray}
I_{\sigma}(\Omega) &=& \int\frac{d\omega}{2\pi}\,\frac{1}{\omega-\frac{\Omega}{2}-E_{{\bf k},-\sigma}+i\Gamma\textrm{sign}\left(\omega-\frac{\Omega}{2}\right)} \quad \\
&& \qquad\times \frac{1}{\omega+\frac{\Omega}{2}-E_{{\bf k},\sigma}+i\Gamma\textrm{sign}\left(\omega+\frac{\Omega}{2}\right)} \ , \nonumber
\end{eqnarray}
$\sigma=\pm 1$ and $E_{{\bf k}\sigma} = k^2/2m + \sigma vk$. This frequency integral can be calculated directly. The imaginary part of $\chi$ yields the interband contribution to the Raman scattering rate:
\begin{eqnarray}\label{Rate3}
\chi''(\Omega) &=& \frac{1}{\pi}\int\frac{d^{3}k}{(2\pi)^{3}}\sum_{\sigma}\gamma_{\sigma,-\sigma}^{\textrm{i,s}}({\bf k})\,\gamma_{-\sigma,\sigma}^{\textrm{s,i}}({\bf k})\,\textrm{Re}\left\lbrace I_{\sigma}^{\phantom{x}}(\Omega)\right\rbrace \nonumber \\
&=& \frac{(mv^{2})^{2}}{2\pi^{4}}I(\hat{{\bf e}}_{\textrm{s}},\hat{{\bf e}}_{\textrm{i}})\int\limits_{0}^{\infty}dk\,k^{2}\sum_{\sigma}\textrm{Re}\left\lbrace I_{\sigma}(|\Omega|)\right\rbrace \ . \quad
\end{eqnarray}
with 
\begin{widetext}
\begin{eqnarray}\label{ReIsigma}
&& 2\pi\,\textrm{Re}\left\lbrace I_{\sigma}(|\Omega|)\right\rbrace = \frac{1}{2}\left(\frac{|\Omega|-2\sigma vk}{(|\Omega|-2\sigma vk)^{2}+4\Gamma^{2}}-\frac{1}{|\Omega|-2\sigma vk}\right)\left\lbrack \ln\left(\frac{E_{{\bf k},\sigma}^{2}+\Gamma^{2}}{(E_{{\bf k},-\sigma}+|\Omega|)^{2}+\Gamma^{2}}\right)+\ln\left(\frac{E_{{\bf k},-\sigma}^{2}+\Gamma^{2}}{(E_{{\bf k},\sigma}-|\Omega|)^{2}+\Gamma^{2}}\right)\right\rbrack \nonumber \\
&&\quad+\frac{2\Gamma}{(|\Omega|-2\sigma vk)^{2}+4\Gamma^{2}}\,\Biggl\lbrack\arctan\left(\frac{E_{{\bf k},-\sigma}+|\Omega|}{\Gamma}\right)-\arctan\left(\frac{E_{{\bf k},\sigma}-|\Omega|}{\Gamma}\right)+\arctan\left(\frac{E_{{\bf k},\sigma}}{\Gamma}\right)-\arctan\left(\frac{E_{{\bf k},-\sigma}}{\Gamma}\right)\Biggr\rbrack \ .
\end{eqnarray}
\end{widetext}

The Raman vertex (\ref{Gamma1}) for Weyl electrons is dominated by the intermediate states $\gamma$ which belong to the nodal spectrum just like the initial $\beta$ and final $\alpha$ states. The intermediate states from higher bands can be handled by the effective mass approximation \cite{Deveraux2007}, but their effect is negligible (formally vanishes) due to the linear relativistic character of the low-energy Weyl spectrum. Focusing on the nodal states only, it will suffice for our purposes to argue that the Raman vertex is approximatelly independent of the momentum magnitude $|{\bf k}|$. First, as usual in condensed matter systems, we may neglect the momentum transfer ${\bf q}$ from the photon to the electrons. Then, all states $\alpha,\beta,\gamma$ live at the same momentum ${\bf k}$, and $\gamma$ can coincide either with $\alpha$ or $\beta$. The ``diamagnetic'' term $\rho_{\alpha\beta}$ in (\ref{Gamma1}) vanishes for interband transitions ($\alpha\neq\beta$). In the other part of (\ref{Gamma1}), one denominator at $\gamma=\alpha$ is equal to the other one at $\gamma=\beta$ and reduced either to just $\omega_{\textrm{i}}$ or $\omega_{\textrm{s}}$ due to energy conservation $\Omega = E_\alpha-E_\beta = \omega_{\textrm{s}}-\omega_{\textrm{i}}$. The corresponding numerators each pick the matrix elements of the gauged momentum operator as $p_{\uparrow\downarrow} \propto \langle\alpha|\boldsymbol{\mathcal{A}}|\beta\rangle$ and $p_{\uparrow\uparrow}, p_{\downarrow\downarrow} \propto \bf k + \langle\alpha|\boldsymbol{\mathcal{A}}|\beta\rangle$; the momentum dependence of the latter cancels out once the two fractions with equal denominators are added up.

The integrals over the direction and magnitude of ${\bf k}$ in (\ref{Rate3}) are separable because $I_\sigma$ depends only on $|{\bf k}|$ and the Raman vertex $\gamma_{\bf k}$ depends (approximatelly) only on $\hat{\bf k}$. Then, the polarization dependence is fully contained in the factor $I(\hat{{\bf e}}_{\textrm{s}},\hat{{\bf e}}_{\textrm{i}})$ constructed from the $\hat{\bf k}$ integral. This factor is not important for our present purpose. The dependence of the Raman scattering rate on the shift frequency $\Omega$ follows only from the remaining momentum-magnitude integral in (\ref{Rate3}), which we calculate numerically and plot in the main text.

\subsection{Quadratic band touching spectrum}

A model Hamiltonian displaying quadratic band touching has been constructed in the context of pyrochlore iridates \cite{Moon2013}
\begin{equation}\label{HamiltonianLuttinger}
H_{\bf k}=\frac{k^{2}}{2\widetilde{M}_{0}}+\frac{\frac{5}{4}k^{2}-({\bf k}{\bf J})^{2}}{2m'}-\frac{(k_{x}J_{x})^{2}\!+\!(k_{y}J_{y})^{2}\!+\!(k_{z}J_{z})^{2}}{2M_{\textrm{c}}}
\end{equation}
It involves three mass parameters $\widetilde{M}_0, M_{\textrm{c}}, m'$ and the $S=\frac{3}{2}$ matrix representations of spin projection operators $J_x, J_y, J_z$. The two-fold degenerate conduction $\sigma=1$ and a valence $\sigma=-1$ bands in the spectrum of this Hamiltonian touch quadratically at ${\bf k}=0$ in a certain domain of mass parameters.

Since the spectrum is quadratic, we can apply the effective mass approximation \cite{Deveraux2007} to calculate the Raman vertex function:
\begin{eqnarray}\label{Gamma4}
\gamma_{{\bf k}}^{\phantom{x}} &\approx& m\,\hat{e}_{\textrm{i}}^{a}\hat{e}_{\textrm{s}}^{b}\frac{\partial^{2}H_{{\bf k}}}{\partial k^{a}\partial k^{b}} \\
&=& m\,\hat{e}_{i}^{a}\hat{e}_{s}^{b}\left(\frac{\delta_{ab}}{M_{0}}+\frac{\frac{5}{2}\delta_{ab}-\lbrace J_{a},J_{b}\rbrace}{2m'}-\frac{J_{a}^{2}\delta_{ab}}{M_{c}}\right) \ . \nonumber
\end{eqnarray}
We can obtain the frequency dependence of the Raman scattering cross-section more easily by neglecting the momentum transfer ${\bf q}$ from the photon to the electrons. In this approximation, the matrix elements $\gamma^{\textrm{i,s}}_{\alpha\beta}$ of the Raman vertex do not depend on the momentum magnitude $|{\bf k}|$. Interband transitions are taken into account by the expression for the scattering rate analogous to (\ref{Rate3}):
\begin{eqnarray}\label{Rate6}
\chi''(\Omega) &=& \frac{1}{\pi}\sum_{nn'}\int\frac{d^{3}k}{(2\pi)^{3}}\sum_{\sigma}\gamma_{\sigma n, -\sigma n'}^{\textrm{i,s}}({\bf k})\,\gamma_{-\sigma n', \sigma n}^{\textrm{s,i}}({\bf k}) \nonumber \\
&& \qquad \times \textrm{Re}\left\lbrace I_{\sigma}^{\phantom{x}}({\bf k},\Omega)\right\rbrace \ ,
\end{eqnarray}
with the main new ingredients being a different spectrum and the summation over a larger set of quantum numbers involving a two-valued index $n$ which distinguishes the degenerate bands. The function $\textrm{Re}\lbrace I_{\sigma} \rbrace$ obtained from the frequency integral is universally given by (\ref{ReIsigma}), but with $2\sigma vk$ systematically replaced by the interband energy difference $E_{{\bf k},\sigma} - E_{{\bf k},-\sigma}$ appropriate for the Luttinger nodal spectrum. The cubic anisotropy complicates calculations without imparting any special qualitative features on Raman scattering. And, if the spectrum is approximated by an isotropic one, we can again separate the integrals over the direction and magnitude of ${\bf k}$. The direction integral is relevant only for the light-polarization dependence of the Raman response, and the momentum integral can be calculated in a straight-forward manner numerically as in the case of Weyl electrons. The resulting dependence of the Raman scattering on the shift frequency $\Omega$ is plotted in the main text. The behavior $\chi''(\Omega)\propto\sqrt{|\Omega|}$ beyond a threshold frequency in the infinite lifetime limit ($\Gamma\to 0$) is physically determined by the electronic density of states.

\subsection{Experimental Methods}

\subsubsection{Raman Scattering Spectroscopy}

Nd$_2$Ir$_2$O$_7$ and Pr$_2$Ir$_2$O$_7$~single crystals were grown by the KF-flux method and possess  as-grown octrahedron-shaped (111) facets. Raman scattering spectra were collected using the Horiba Jobin-Yvon T64000 triple monochromator spectrometer with a liquid nitrogen cooled CCD detector. Our measurements were performed in pseudo-Brewster's angle geometry with a \SI{50}{\micro\meter} laser probe diameter. The 514.5 nm line of Ar$^+$-Kr$^+$ mixed gas laser was used as the excitation light. The intensity of the incident light was 15 mW, and the laser heating was estimated to be 15 K. The samples were mounted on the cold-finger of Janis ST-500 cryostat, which can be cooled down to 4 K without laser heating.  

The polarization-resolved spectra were measured in four configurations: $\hat{z}(xx)z$, $\hat{z}(xy)z$, $\hat{z}(RR)z$, and $\hat{z}(RL)z$, where $R$($L$) denotes the right (left) circular polarization.  In the main manuscript, we were using data obtained in the $\hat{z}(xy)z$ polarization in order to separate the electronic Raman scattering we are interested in from the $A_{1g}$ magnetic scattering.

The intensity of all spectra was normalized by the thermal factor $[n(\omega, T)+1]$, where $n(\omega, T)$ is the Bose occupation factor.

To correct for  the small deviations (less than 20 \%  for the same excitation power)  of the intensity of Raman response in different measurements, all the spectra were normalized to the intensity of the $A_{1g}$ phonon. 

The electronic Raman continuum spectra presented in the main text Fig.~2 and Fig.~3 were obtained by subtracting the phonon and magnon features from the original spectra, which will be published elswhere~\cite{Xu2022prep}. In order to do that, the original spectra were fitted with a number of Lorentzian shapes, and those relevant to phonons and magnons were subtracted from the spectra.

\subsubsection{Disorder Characterization}

\begin{figure}
\centering
\includegraphics[width=0.45\textwidth]{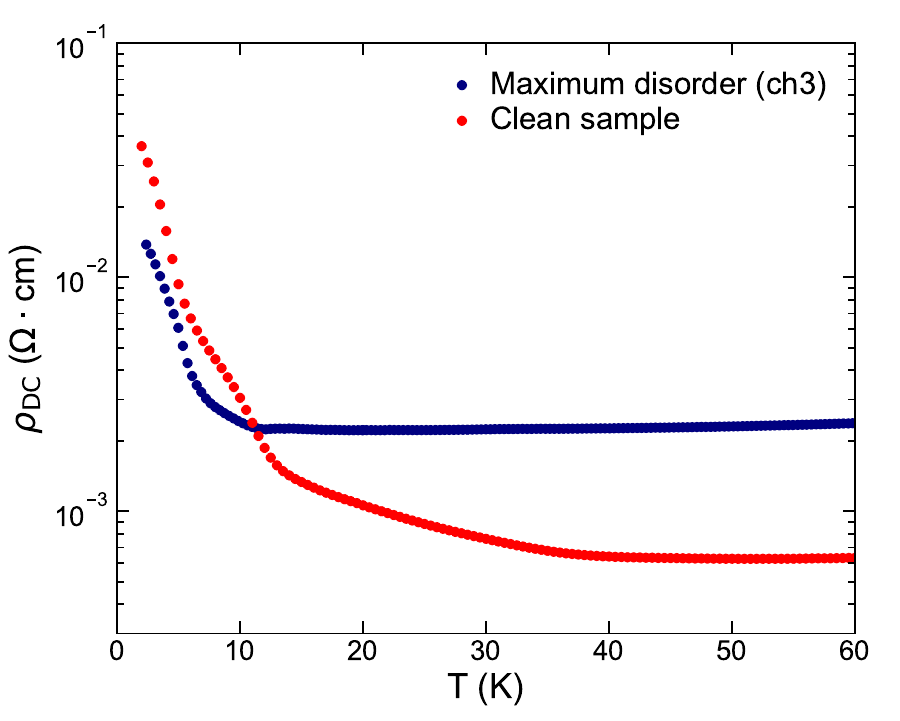}
\caption{Temperature dependence of the resistivity of the ``clean" and ``maximum disorder" (ch3) Nd$_2$Ir$_2$O$_7$ crystals used in the Raman scattering experiment. }
\label{nio_rho}
\end{figure}

In Fig.~3 of the main manuscript we present the data for three different crystals of Nd$_2$Ir$_2$O$_7$  with different amounts of disorder. It is known that disorder in  Nd$_2$Ir$_2$O$_7$ can be characterized by resistivity measurements, where disorder was shown to introduce a metallic behaviour of resistivity~\cite{Ueda2012}. Resistivity measurements presented in Fig.~\ref{nio_rho}  demonstrate an increase of disorder from the ``clean sample'' to the ``maximum disorder'' ch3 sample.

\begin{figure}
\centering
\includegraphics[width=0.45\textwidth]{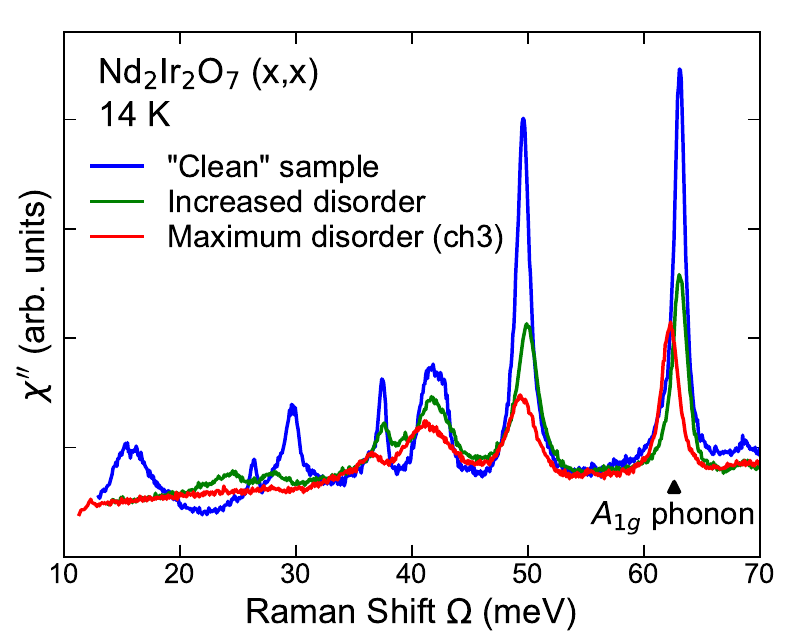}
\caption{Raman scattering spectra at 14 K including phonons and magnon excitations  of three crystals of  Nd$_2$Ir$_2$O$_7$ with different disorder levels in the $(x,x)$ polarization. We observe an apparent broadening of the phonons on the increase of the levels of structural disorder.  }
\label{nio_phonons}
\end{figure}

Structural disorder can be also characterized by the width of the phonons, which become broader as structural disorder increases~\cite{Drichko2016}. Fig.~\ref{nio_phonons} compares the phonons of the three Nd$_2$Ir$_2$O$_7$ crystals and visually demonstrates the broadening of all the phonons of oxygen vibrations observed above 35 meV~\cite{Xu2022}, in addition to dramatic changes of magnons, observed below 30 meV. For example, the width of a non-degenerate $A_{1g}$ phonon at 63 meV which cannot be affected by symmetry breaking, but is affected by disorder, increases from 1.15 meV in the ``clean'' sample to 1.43~meV in the ``increased disorder'' sample, to 1.57 meV in ch3.

\end{document}